\documentclass{article}
\usepackage{graphicx}
\usepackage{amsmath}
\usepackage{amsfonts}
\usepackage{amssymb}

\begin{document}

\title{Non-holonomic Control II : Non-holonomic Quantum Devices}
\author{E. Brion\\\emph{Laboratoire Aim\'{e} Cotton, }\\\emph{CNRS II, B\^{a}timent 505, }\\\emph{91405 Orsay Cedex, France.}
\and V.M. Akulin\\\emph{Laboratoire Aim\'{e} Cotton, }\\\emph{CNRS II, B\^{a}timent 505, }\\\emph{91405 Orsay Cedex, France.}
\and D. Comparat\\\emph{Laboratoire Aim\'{e} Cotton, }\\\emph{CNRS II, B\^{a}timent 505, }\\\emph{91405 Orsay Cedex, France.}
\and I. Dumer\\\emph{College of Engineering, }\\\emph{University of California, }\\\emph{Riverside, CA 92521, USA. }
\and V. Gershkovich\\\emph{Institut des Hautes Etudes Scientifiques,}\\\emph{ Bures-sur-Yvette, France. }
\and G. Harel\\\emph{Department of Computing, }\\\emph{University of Bradford, }\\\emph{Bradford, West Yorkshire BD7 1DP, United Kingdom. }
\and G. Kurizki\\\emph{Department of Chemical Physics, }\\\emph{Weizmann Institute of Science, }\\\emph{76100 Rehovot, Israel. }
\and I. Mazets\\\emph{Department of Chemical Physics techni, }\\\emph{Weizmann Institute of Science, }\\\emph{76100 Rehovot, Israel. }\\\emph{A.F. Ioffe Physico-Technical Institute, }\\\emph{194021 St. Petersburg, Russia. }
\and P. Pillet\\\emph{Laboratoire Aim\'{e} Cotton, }\\\emph{CNRS II, B\^{a}timent 505, }\\\emph{91405 Orsay Cedex, France.}}
\maketitle
\begin{abstract}
In this paper, we show how the non-holonomic control technique can be employed to build completely controlled quantum devices. Examples of such controlled structures are provided.   
\end{abstract}

\section{Introduction}
In the previous article, we showed that quantum systems become ``non-holonomic'' when perturbed in a certain
time-dependent way : as a result of the perturbation, all global constraints on the dynamics are removed and the
system becomes fully controlled. The straightforward application of the non-holonomic control to groups of interacting two-level systems, the so-called qubits, is a promising way to achieve computational algorithms widely discussed in quantum informatics \cite{Feynman:82, Lloyd:93, DiVincenzo:95, Cirac:95}. Indeed, quantum computations \cite{Ekert:96} are achieved through combining well-chosen quantum gates \cite{Tgate}, which are the analog of the classical logic gates and consist in particular evolutions of the system chosen as a computer. In this context, the non-holonomic control is a precious means for implementing any gate and thus performing any computation on an arbitrary quantum system. However, the direct control of an $N$-qubit computer's evolution is a very heavy (or even intractable) computational task when $N$ is a large number, since it requires the control of $4^{N}$ physical parameters. To overcome this impediment one can combine completely controlled cells in a manner which depends on the kind of the problem to be solved : the adaptation of the structure of the quantum device, obtained as a particular arrangement of controlled elements, to the computation to be performed allows one to decrease the number of the free control parameters needed, and thus the complexity of the control problem. The complete control of the whole compound device is thus assured by the controllability of
the individual cells as well as their connections with each other within the
architecture of the computer.

In the first section of this paper, we provide an example of a unit cell, which is completely controlled
through non-holonomic interactions. In the second section, we propose two different devices
composed of such cells, the arrangements of which suit particularly well
universal quantum computations and simulation of quantum field dynamics,
respectively. In the third section, we finally describe a toy device that can perform quantum
computations on 9 qubits and show in particular how it can perform the
discrete Fourier transform on 9 qubits.

\section{Completely Controlled Unit Cell}

One way to construct a completely controlled but not
immediately universal quantum device is to build it up from small parts, called
``unit cells'', each of which is non-holonomic and therefore directly and
universally controllable. The proper functioning of the device relies then on
the appropriate connection of the cells. In this way the universality of the
device is obtained indirectly, not by applying a huge number of controls, but
by smartly connecting the cells and choosing the sequence of operations
performed. There is no general prescription on how to construct a particular
device; this requires expertise in the art of ``programming'' the operations
of the cells and their interactions.

\subsection{Cell structure}

\begin{figure}
[ptbh]
\begin{center}
\includegraphics[
height=2.9412in,
width=3.1678in,
angle=-90
]%
{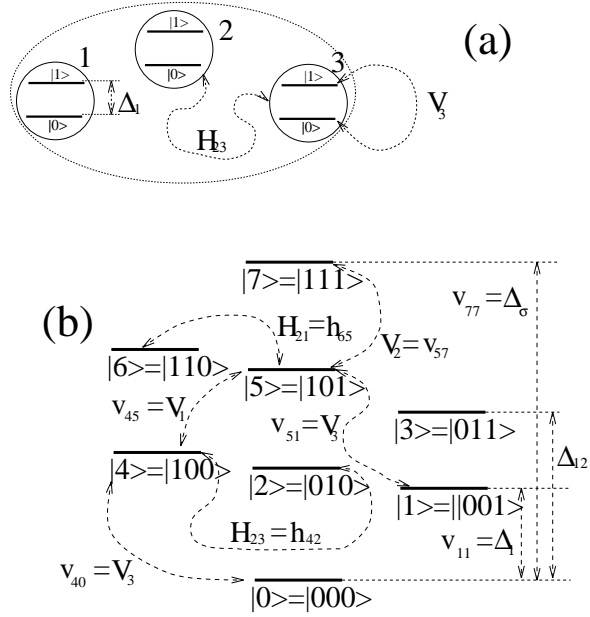}%
\caption{Realization of a unit cell: A compound system of three two-level
atoms interacting with external electromagnetic and static electric fields.
(a) The $i$th atom has ground and excited states $|0\rangle_{i}$ and
$|1\rangle_{i}$ with excitation energy $A_{i}+\Delta_{i}$ that can be modified
by the static field; transition amplitude in the electromagnetic field is
$V_{i}$; the dipole-dipole coupling of the $i$th and $j$th atoms is $D_{ij}$.
(b) The computational basis states and their relation to matrix elements of
the principal Hamiltonian $\widehat{H}_{0}$ and the perturbations $\widehat{P}%
_{\omega}$ and $\widehat{P}_{S}$ of Eqs.~(\ref{eq:one}-\ref{eq:onevs}).}%
\label{Fig3}%
\end{center}
\end{figure}

Fig.~\ref{Fig3} shows an example of a completely controlled unit cell composed of three two-level atoms, each with ground and excited states $|0\rangle$ and $|1\rangle$, having distinct transition
frequencies $\omega_{1}^{a}$, $\omega_{2}^{a}$, and $\omega_{3}^{a}$. The
atoms are subject to dipole-dipole interactions and are coupled to two
external fields: an electromagnetic field $E_{\omega}=\mathcal{E}_{\omega
}\mathrm{cos}\omega t$ of nearly resonant frequency $\omega$, and a static
electric field $E_{S}$. The dipole-dipole interaction is fixed and determines
the principal, unperturbed Hamiltonian of the system, $\widehat{H}_{0}$, while the
external fields provide two controllable perturbations, $\widehat{P}_{\omega}$ and
$\widehat{P}_{S}$. The Hilbert space of the system has a ``computational basis''
of $N=2^{3}=8$ states, $|x\rangle\equiv|x_{2}x_{1}x_{0}\rangle\equiv
|x_{2}\rangle|x_{1}\rangle|x_{0}\rangle$, $x=0,1,\dots,7$, where the state of
the $i$th atom encodes the $i$th binary digit of $x=\sum_{r=0}^{2}x_{r}2^{r}$
as a qubit [see Fig.~\ref{Fig3}(b)]. The crucial requirement is the
non-holonomic character of the interaction : $\widehat{H}_{0}$,
$\widehat{P}_{\omega}$, $\widehat{P}_{S}$, and their commutators of all orders must span
the linear space of $8\times8$ Hermitian matrices. This is indeed the case for
the system shown in Fig.~\ref{Fig3}, for which principal Hamiltonian and
perturbations are given, in the computational basis and assuming resonant
approximation, by the matrices
\begin{align}
\widehat{H}_{0}  &  =\left(
\begin{array}
[c]{cccccccc}%
0 & 0 & 0 & 0 & 0 & 0 & 0 & 0\\
0 & A_{1} & D_{12} & 0 & D_{13} & 0 & 0 & 0\\
0 & D_{21} & A_{2} & 0 & D_{23} & 0 & 0 & 0\\
0 & 0 & 0 & A_{12} & 0 & D_{23} & D_{13} & 0\\
0 & D_{31} & D_{32} & 0 & A_{3} & 0 & 0 & 0\\
0 & 0 & 0 & D_{32} & 0 & A_{13} & D_{21} & 0\\
0 & 0 & 0 & D_{31} & 0 & D_{12} & A_{23} & 0\\
0 & 0 & 0 & 0 & 0 & 0 & 0 & A_{\sigma}%
\end{array}
\right)  ,\label{eq:one}\\
C_{\omega}\widehat{P}_{\omega}  &  =\left(
\begin{array}
[c]{cccccccc}%
0 & V_{1} & V_{2} & 0 & V_{3} & 0 & 0 & 0\\
V_{1} & 0 & 0 & V_{2} & 0 & V_{3} & 0 & 0\\
V_{2} & 0 & 0 & V_{1} & 0 & 0 & V_{3} & 0\\
0 & V_{2} & V_{1} & 0 & 0 & 0 & 0 & V_{3}\\
V_{3} & 0 & 0 & 0 & 0 & V_{1} & V_{2} & 0\\
0 & V_{3} & 0 & 0 & V_{1} & 0 & 0 & V_{2}\\
0 & 0 & V_{3} & 0 & V_{2} & 0 & 0 & V_{1}\\
0 & 0 & 0 & V_{3} & 0 & V_{2} & V_{1} & 0
\end{array}
\right)  ,\label{eq:onevem}\\
C_{S}\widehat{P}_{S}  &  =\left(
\begin{array}
[c]{cccccccc}%
0 & 0 & 0 & 0 & 0 & 0 & 0 & 0\\
0 & \Delta_{1} & 0 & 0 & 0 & 0 & 0 & 0\\
0 & 0 & \Delta_{2} & 0 & 0 & 0 & 0 & 0\\
0 & 0 & 0 & \Delta_{12} & 0 & 0 & 0 & 0\\
0 & 0 & 0 & 0 & \Delta_{3} & 0 & 0 & 0\\
0 & 0 & 0 & 0 & 0 & \Delta_{13} & 0 & 0\\
0 & 0 & 0 & 0 & 0 & 0 & \Delta_{23} & 0\\
0 & 0 & 0 & 0 & 0 & 0 & 0 & \Delta_{\sigma}%
\end{array}
\right)  . \label{eq:onevs}%
\end{align}
Here $D_{ij}=d_{i}d_{j}/R_{ij}^{3}$ denotes the dipole-dipole coupling of the $i$th
and $j$th atoms at distance $R_{ij}$, with $d_{i}$ the $i$th atom dipole
matrix element, and $V_{i}=\mathcal{E}_{\omega}d_{i}$ is the dipole coupling
of the $i$th atom to the external electromagnetic field. The excitation
energy detunings of single atoms $A_{i}=\hbar(\omega_{i}^{a}-\omega)$
determine the detunings of pairs of atoms $A_{ij}=A_{i}+A_{j}$ and the total
detuning $A_{\sigma}=A_{1}+A_{2}+A_{3}$. Their values can be changed by
variation of a static electric field ${E_{S}}$ (Stark effect), which results
in energy shifts $A_{i}\rightarrow A_{i}+\Delta_{i}$ for single atoms, where
$\Delta_{i}=\alpha_{i}E_{S}$ depend on atom-specific electric permeability
constants $\alpha_{i}$, and similar shifts $\Delta_{ij}=\Delta_{i}+\Delta_{j}$
and $\Delta_{\sigma}=\Delta_{1}+\Delta_{2}+\Delta_{3}$ for two and three
atomic detunings respectively.%
%EndExpansion

Note that, by a proper choice of $\Delta_{i}$ and $\omega$, one can set two of
three $A_{i}$ to zero. Moreover, for clarity we also
set to zero the third $A_{i}$, which remains just a part of $\widehat{H}_{0}$
otherwise. Hence, hereafter, all $\Delta_{i}$ denote just the deviations from
zero resulting from the variation ${E_{S}}$ of the Stark field. The last
together with the amplitude $=\mathcal{E}_{\omega}$ serve as a time dependent
control parameters $C_{S}$ and $C_{\omega}$ respectively. The matrices $\widehat
{P}_{S}$ and $\widehat{P}_{\omega}$ contain therefore only the permeabilities
$\alpha_{i}$ and the dipole moments $d_{i}$ respectively.

\subsection{Cell control}

To exert direct universal control over the unit cell we employ the non-holonomic control technique presented in the previous article. (i) We fix $N^{2}=64$ consecutive time intervals of equal length $\tau=\frac{T}{64}$, during which the two perturbations are alternately applied to the system : in the $k$th interval the perturbation is $\widehat{P}_{k}=\widehat{P}_{S}$ for odd $k$ and $\widehat{P}_{k}=\widehat{P}_{\omega}$ for even $k$, where $k=1,2,\dots,64$. The strength of $\widehat{P}_{k}$ is denoted by $C_{k}$, and corresponds either to $\mathcal{E}_{\omega}$ or $E_{S}$, depending on the parity of $k$. Thus, the evolution of the system is governed by a Hamiltonian which is constant on each interval:
\begin{equation}
\widehat{H}(t)=\widehat{H}_{0}+C_{k}\widehat{P}_{k}\ \ \ \ \ \ t\in\lbrack(k-1)\tau,k\tau].
\end{equation}
(ii) We look for the $64$-dimensional vector $\overrightarrow{C}^{(0)}$ such that
\begin{equation}
\widehat{U}\left( \overrightarrow{C}^{(0)} \right)\equiv\prod_{k=1}^{64}\exp\left[  -\frac{i}{\hbar}(\widehat{H}_{0}+C_{k}^{(0)}\widehat{P}_{k})\tau \right]  =\widehat{I}. \label{eq:unity}%
\end{equation}
To this end, we first solve the ``$8$th root'' of Eq.~(\ref{eq:unity}),
\begin{equation}
\prod_{k=1}^{8}\exp\left[  -\frac{i}{\hbar}(\widehat{H}_{0}+c_{k}\widehat{P}_{k})\tau\right]  =\widehat{I}^{1/8}, \label{eq:nrootunity}
\end{equation}
by minimizing the functional $\sum_{j=0}^{8}\left|  a_{j}\left(
\{c_{k}\}_{k=1\ldots 8}\right)  \right|  ^{2}$ to $2$, where $\left\{  a_{j}\right\}
$ denote the coefficients of the characteristic polynomial of the matrix
product in Eq.(\ref{eq:nrootunity}). This provides a sequence of eight values, $c_{1},c_{2},\dots,c_{8}$, the repetition of which yields the desired vector $\overrightarrow{C}^{(0)}$. (iii) Finally, we compute the required vector $\overrightarrow{C}$ as described in the previous article. If the target evolution $\widehat{U}_{arbitrary}=\widehat{U}_{\epsilon}\equiv\exp(-i\widehat{\mathcal{H}}\epsilon)$ is close to the identity (\emph{i.e.} $\epsilon$ is small), one determines the variations $\delta
C_{k}$ to first order in $\epsilon$ by solving the linear equations
\begin{equation}
\sum_{k=1}^{64}\frac{\partial\widehat{U}}{\partial C_{k}}\left( \overrightarrow{C}^{(0)} \right)\,\delta
C_{k}=-i\widehat{\mathcal{H}}\epsilon,
\end{equation}
and, replacing $\overrightarrow{C}^{(0)}$ by $\overrightarrow{C}^{(0)}+ \delta \overrightarrow{C}$, one repeats the same operation, and so on, until one gets the vector $\overrightarrow{C}$ which checks $\widehat{U} \left(\overrightarrow{C}\right) = \widehat{U}_{\epsilon}$ with desired accuracy. To perform an arbitrary unitary
transformation $\widehat{U}_{arbitrary}=\widehat{U}_{\epsilon}$, with $\epsilon$ taking any value in $[0,2\pi]$ and not necessarily small, we divide the work into ``small'' steps : we apply the transformation $\widehat{U} \left(\overrightarrow{C}^{\left(\frac{1}{n^{\ast}}\right)}\right)=\widehat{U}_{\epsilon/n^{\ast}}=\left(\widehat{U}_{arbitrary}\right)^{\frac{1}{n^{\ast}}}$ repeatedly $n^{\ast}$ times, with $n^{\ast}$ determined as described in the previous article, and obtain
\begin{equation}
\left[\widehat{U}\left(\overrightarrow{C}^{\left(\frac{1}{n^{\ast}}\right)}\right)\right]^{n^{\ast}}=\left[\widehat{U}_{\epsilon/n^{\ast}}\right]^{n^{\ast}}= \left[\left(\widehat{U}_{arbitrary}\right)^{\frac{1}{n^{\ast}}}\right]^{n^{\ast}} = \widehat{U}_{arbitrary}.
\end{equation}

\begin{figure}
[ptbh]
\begin{center}
\includegraphics[
height=3.8735in,
width=3.7213in,
angle=-90
]
{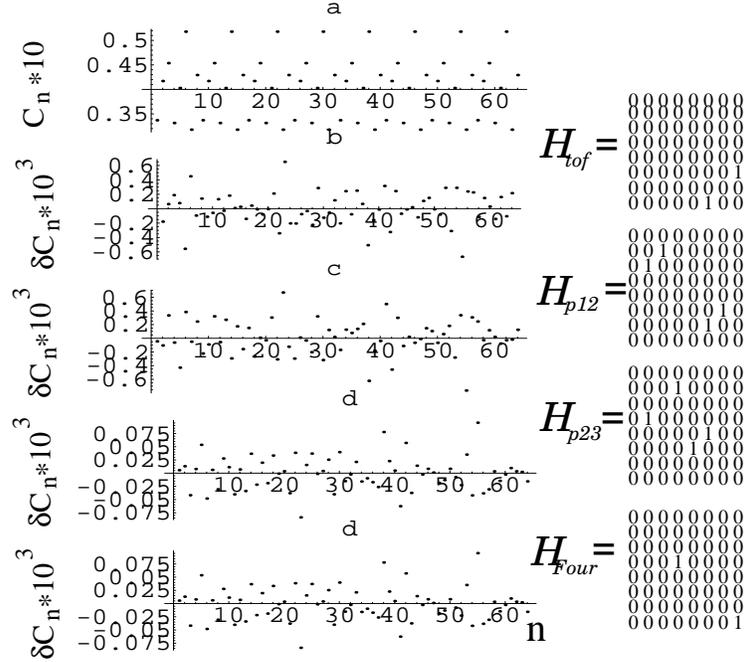}
\caption{(a) Control parameters $C_{k}^{0}$ for the identity transformation
$\widehat{I}$. Variations $\delta C_{k}$ achieving the transformation
$\widehat{U}_{\epsilon/8}$ on the cell, with $\widehat{U}_{\epsilon}\equiv(\widehat{U}_{\epsilon
/8})^{8}$ equal to: (b) the permutation $\widehat{U}_{p12}$; (c) the permutation
$\widehat{U}_{p23}$; (d) the Toffoli-gate transformation $\widehat{U}_{Toff}$. (e)
Variations $\delta C_{k}$ achieving the conditional phase shift $\widehat{B}(\phi)=\exp(-i\phi\widehat{\mathcal{H}}_{B})$, at $\phi=\pi/32$, employed in the
quantum discrete Fourier transform.}
\label{Fig4}%
\end{center}
\end{figure}

In Fig. \ref{Fig4} we show examples of unit cell control, where
appropriately chosen parameters $C_{k}^{(0)}$ and variations $\delta C_{k}$ achieve
unitary transformations on the unit cell: the Toffoli-gate transformation (see
Appendix~\ref{appendix-a}), two-qubit permutations $\widehat{p}_{ij}|a\rangle
_{i}|b\rangle_{j}=|b\rangle_{i}|a\rangle_{j}$ $(a,b=0,1)$, and the conditional
phase shift employed in the quantum discrete Fourier transform (discussed in
Sec.~\ref{toy}). The transformation is achieved either directly $(n^{\ast}=1)$ or by
8 repetitions $(n^{\ast}=8)$. The operators $\widehat{H}_{0}$, $\widehat{P}_{\omega}$ and
$\widehat{P}_{S}$ are chosen with arbitrary realistic values. We take
$D_{12}=1.1E_{u}$, $D_{23}=0.946E_{u}$, $D_{13}=0.86E_{u}$, and $T=250\hbar
/E_{u}$, where $E_{u}\sim10^{-18}$ erg is the typical energy scale. For odd
$k$ we switch off the external electromagnetic field, $V_{1;2;3}=0$, and tune
the atomic excitation energies by the Stark field $E_{S}$ such that
$\Delta_{1;2;3}=(0.1;0.11;0.312)E_{u}$. For even $k$ we set $E_{S}=0$, that is
$\Delta_{1;2;3}=0$, and take $V_{1;2;3}=(0.3;0.33;0.24)E_{u}$.%

\section{Completely Controlled Quantum Devices}

\label{devices}
\begin{figure}
[ptbh]
\begin{center}
\includegraphics[
height=4.2194in,
width=5.2667in,
angle=-90
]%
{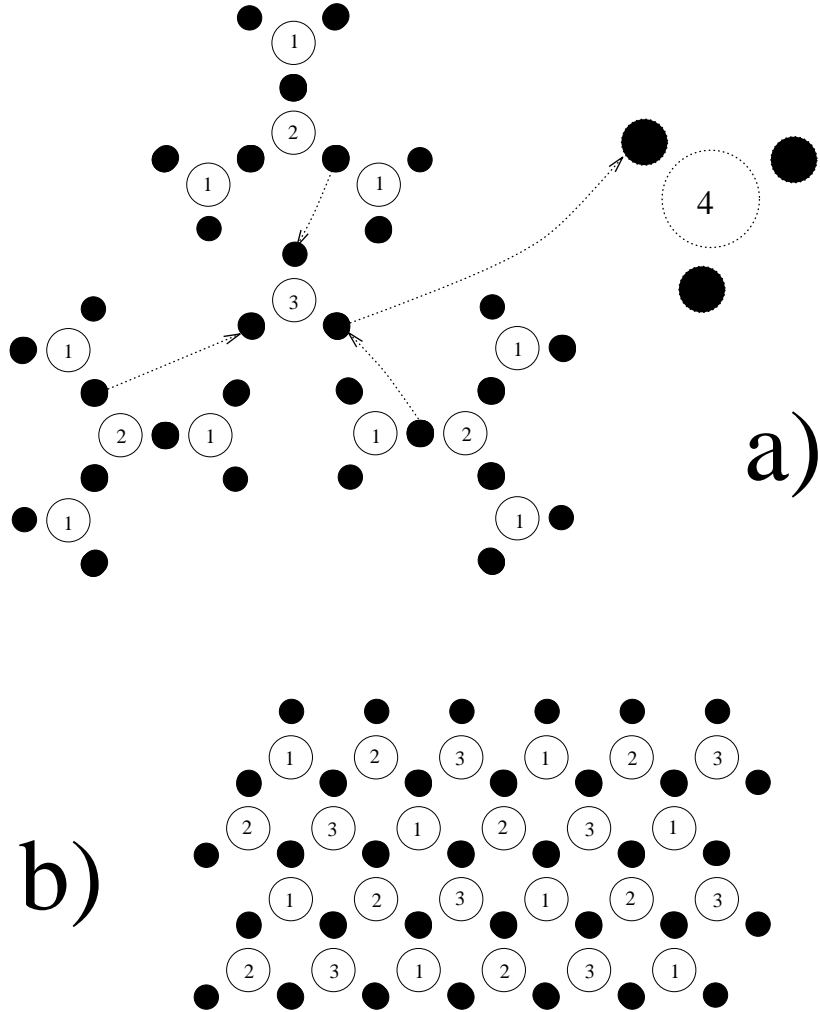}%
\caption{Two possible arrangements of cells for special purpose devices: (a)
tree-like structure for quantum computation; (b) planar lattice for simulating
dynamics of quantum fields. The circled numbers denote the rank of joints of
the tree (a) or specify the order in which atoms are grouped into triads (b).
The arrows show state exchange to parent joints.}%
\label{Fig5}%
\end{center}
\end{figure}

Once completely controlled unit cells have been constructed, a compound device can be assembled from such elements in an architecture which depends on the specific function it has to perform. Fig. \ref{Fig5} shows two possible arrangements of unit cells designed for two different purposes : the first one suites more the purpose of quantum computing, while the second is more useful for simulating lattice quantum field dynamics.

The first device (Fig.~\ref{Fig5}(a)) is organized in a tree-like
structure. In this arrangement, the quantum state of one atom in each cell can be exchanged
with the state of an atom in the closest parent joint of the tree. Even though the
simplest way to make the exchange is to straightforwardly displace the atom to the parent joint, the exchange or transport of the state without moving the atom can be
more practical. The tree-like architecture and the possibility to perform all
the unitary transformations (including all the permutations) in each unit cell
allow one to put together the states of any three
two-level atoms of the device and make them interfere after at most $s=6\log_{3}n$ state exchanges, by
moving them toward the root of the tree to a common cell. Placing the new
states back (if needed) requires the same number of inverse exchanges. This is
a very modest number, $s\sim40$, even for a rather large device of
$n\sim10^{3}$ with Hilbert space of $N=2^{n}\sim10^{300}$ dimensions. Hence,
all basic operations of quantum computation can be performed on any physical
system composed of non-holonomic triads of two-level subsystems arranged in a
tree-like structure, and each operation can be completed within $64\times
16\times12\times\log_{3}n$ control intervals $\tau$. Note that the identity
transformation should be applied to all other cells to preserve their states
during the operation.
%EndExpansion

The second arrangement of cells (Fig.~\ref{Fig5}(b)) is mainly designed to emulate the dynamics of quantum fields on lattices. Of course, it can
also perform general operations on any triad, but for a higher cost of
$s=O(n^{1/2})$. In this arrangement, after each control period of $64 \tau$ the
closest neighboring atoms are differently regrouped in triads (cells), with
the original grouping repeating itself after three consecutive periods.
Therefore, at each moment the change of the cell state depends on the states
of the neighboring cells, as it should be in order to emulate the dynamics of
the fields. Immediate analogy to the Ising model emerges when we restrict
ourselves to small values of $\epsilon$ for which terms of order $\epsilon^{2}$
are negligible, and then each $T=64\tau$ period plays the role of the time
increment $\Delta\tau=\epsilon$. The evolution of such a device is determined by
three sums of effective cell Hamiltonians, $\widehat{H}_{eff}^{(p)}=\sum_{q}%
\widehat{\mathcal{H}}_{q,p}$, one for each period $p=1,2,3$, where $\widehat
{\mathcal{H}}_{q,p}$ is the effective Hamiltonian of the $q$th cell at the
$p$th period.

We can cast the cell Hamiltonians to sums of tensor products of Pauli matrices
$\widehat\sigma^{i}_{\alpha}$, where the Greek index $\alpha=x,y,z$ denotes the
matrix type and the Latin index $i$ specifies the two-level atom on which it
acts. Since the cells are under complete control, the coefficients of this
development can be made an arbitrary function of the time $\tau$, and hence
the effective Hamiltonian reads
\begin{align}
\widehat H_{eff}(\tau)  &  = A_{i}^{\alpha}(\tau) \widehat\sigma_{\alpha}^{i}
+B_{(i,j)}^{\alpha\beta}(\tau) \widehat\sigma_{\alpha}^{i} \widehat\sigma_{\beta}%
^{j}\nonumber\\
&  +\,C_{(i,j,k)}^{\alpha\beta\gamma}(\tau) \widehat\sigma_{\alpha}^{i} \widehat
\sigma_{\beta}^{j} \widehat\sigma_{\gamma}^{k} ,
\end{align}
with implicit summation over repeated indices, where $(i,j)$ and $(i,j,k)$
indicate pairs and triads of distinct atoms that are periodically grouped in a
common cell. This Hamiltonian results in the evolution equation for the
Heisenberg operators $\widehat\sigma_{\alpha}^{i}(\tau)$,
\begin{align}
\hbar\frac{d\widehat\sigma_{\alpha}^{i}(\tau)}{d\tau}  &  = \mathcal{A}
_{\alpha,j}^{i,\beta} (\tau) \widehat\sigma_{\beta}^{j}(\tau) +\mathcal{B}%
_{\alpha,(j,k)}^{i,\beta\gamma}(\tau) \widehat\sigma_{\beta}^{j}(\tau)\widehat
\sigma_{\gamma}^{k}(\tau)\nonumber\\
&  +\, \mathcal{C}_{\alpha,(j,k,l)}^{i,\beta\gamma\delta}(\tau) \widehat
\sigma_{\beta}^{j}(\tau)\widehat\sigma_{\gamma}^{k}(\tau)\widehat\sigma_{\delta}%
^{l}(\tau), \label{eq:two}%
\end{align}
where the coefficients $\mathcal{A,B,C}$ are determined by $A,B,C$ and the
commutation relations of the Pauli matrices. By a proper choice of the
coefficients $A,B,C$ through the appropriate control sequences, one can
simulate different linear and non-linear lattice models of quantum fields with
time dependent parameters.

\section{Toy Device}

\label{toy} 
\begin{figure}
[ptbh]
\begin{center}
\includegraphics[
height=4.2211in,
width=3.5051in,
angle=-90
]%
{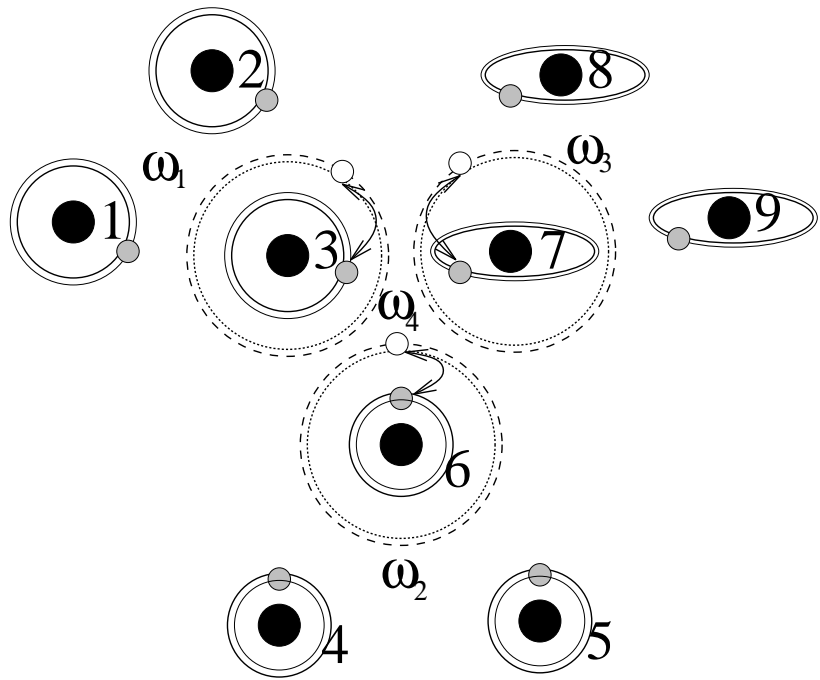}%
\caption{A toy device, composed of $9$ Rydberg atoms, which can perform quantum
computations on 9 qubits. Each atom is a two-level system shown schematically
by double orbits. Atoms of different triads are excited to distinct pairs of
Rydberg states. Each triad $p$ is controlled by an external field of distinct
frequency $\omega_{p}$. One atom in each triad can be excited to a pair of
higher Rydberg states, thus forming a higher-level triad: $(3,6,7)$. These
excitations (depicted by arrows) correspond to state transportations. }%
\label{Fig6}%
\end{center}
\end{figure}

To conclude this paper, we now describe a toy device that can perform quantum computations
on 9 qubits. An ensemble of 9 different Rydberg atoms, \emph{i.e.} atoms of different elements or identical atoms
which are excited to distinct pairs of Rydberg states, is placed in a
magneto-optical trap at low temperature, as illustrated in Fig. \ref{Fig6}. The best candidates for
such a device are the long-living states corresponding to large angular
momentum. By placing all the atoms in a static electric field, one lifts the
degeneracy in the magnetic quantum number and performs tuning if needed. All
the atoms experience the dipole-dipole interaction $\widehat{D}_{ij}=\widehat{d}%
_{i}\widehat{d}_{j}\langle R_{ij}^{-3}\rangle$, where the cube of the inverse
distance between atoms is averaged over their translational quantum states.
Note, however, that only for almost resonant atoms this interaction is
important. By a proper choice of the atomic states and the static field
${E_{S}}$, we obtain three triads, $p=1,2,3$, each of which is composed of three almost
resonant two-level atoms with transition frequencies centered on distinct frequencies $\omega_{p}$. For each triad $p$, the interactions $\widehat
{D}_{ij}$ give the principal Hamiltonian, while a microwave field
$E_{\omega_{p}}$ at the frequency $\omega_{p}$ serves as a control
perturbation. Transportation of the state of one atom in each triad to the
parent joint can be performed by dipole or Raman $\pi$ transitions from the
initial pair of Rydberg levels to a higher pair. With these higher pairs
assumed nearly resonant with a frequency $\omega_{4}$, atoms 3, 6 and 7 form a
higher-level triad---the parent joint of the first three triads---which is
controlled by a fourth microwave field $E_{\omega_{4}}$ of frequency
$\omega_{4}$.
%EndExpansion

As an application of our technique, we show how to perform the discrete Fourier
transform modulo $N=2^{9}=512$ on the toy device we have just presented. This operation is achieved by the following $9$-qubit unitary transformation
\begin{equation}
\widehat{F}_{N}|x\rangle=\frac{1}{\sqrt{N}}\sum_{y=0}^{N-1}\exp(2\pi
ixy/N)|y\rangle, \label{DFTN}%
\end{equation}
where $|x\rangle$ and $|y\rangle$ belong to the system computational
basis, the states of which are defined by
\begin{equation}
|x\rangle\equiv|x_{8}\rangle_{9}\dots|x_{1}\rangle_{2}|x_{0}\rangle_{1},
\end{equation}
with $x\equiv\sum_{r=0}^{8}x_{r}2^{r}=0,1,\dots,N-1$ $(x_{r}=0,1)$, and where
$|\ \rangle_{i}$ denotes the state of the $i$th atom---the $i$th qubit. The
algorithm we employ to perform the Fourier transform is based on constructing
the exponent in Eq.~(\ref{DFTN}) as
\begin{equation}
\exp(2\pi ixy/2^{9})=\prod_{r=0}^{8}\prod_{s=0}^{r}\exp(i\pi x_{r}^{\prime}y_{s}/2^{r-s}),
\end{equation}
where $x_{r}^{\prime}\equiv x_{8-r}$. We begin by reversing the order in which
the bits of the input $x$ are stored in our 9-qubit register, that is, we
achieve the unitary transformation
\begin{equation}
|x_{8}\rangle_{9}\dots|x_{1}\rangle_{2}|x_{0}\rangle_{1}\ \rightarrow
\ \ |x_{0}\rangle_{9}\dots|x_{7}\rangle_{2}|x_{8}\rangle_{1} \label{DFTrev}%
\end{equation}
by applying a sequence of state exchanges. Then we complete the transform in the following 9
steps: (i) We ``split'' the first qubit (the state of atom 1) by applying the
unitary transformation
\begin{equation}
\widehat{A}\equiv\frac{1}{\sqrt{2}}\left(
\begin{array}
[c]{cc}%
1 & 1\\
1 & -1
\end{array}
\right)  =\exp\left[  \frac{-i\pi}{\sqrt{8}}\left(
\begin{array}
[c]{cc}%
1-\sqrt{2} & 1\\
1 & -1-\sqrt{2}%
\end{array}
\right)  \right]  ,
\end{equation}
which maps $|0\rangle\rightarrow\frac{1}{\sqrt{2}}(|0\rangle+|1\rangle)$ and
$|1\rangle\rightarrow\frac{1}{\sqrt{2}}(|0\rangle-|1\rangle)$. Note that this
would already complete the Fourier transform if we had only one qubit. (ii)
Next, we apply to the first and second qubits the conditional phase shift
$|a\rangle_{2}|b\rangle_{1}\rightarrow e^{i\pi ab/2}|a\rangle_{2}|b\rangle
_{1}$ $(a,b=0,1)$, explicitly given by
\begin{equation}
\widehat{B}_{21}\equiv\left(
\begin{array}
[c]{cccc}%
1 & 0 & 0 & 0\\
0 & 1 & 0 & 0\\
0 & 0 & 1 & 0\\
0 & 0 & 0 & e^{i\pi/2}%
\end{array}
\right)  =\widehat{B}(\pi/2),
\end{equation}
where $\widehat{B}(\phi)$ is the unitary transformation
\begin{equation}
\widehat{B}(\phi)=\exp\left[  -i\phi\left(
\begin{array}
[c]{cccc}%
0 & 0 & 0 & 0\\
0 & 0 & 0 & 0\\
0 & 0 & 0 & 0\\
0 & 0 & 0 & -1
\end{array}
\right)  \right]  .
\end{equation}
Then we ``split'' the second qubit by applying the transformation
$\widehat{A}$. This accounts for the contribution of the second most significant
bit of the input $x$. (iii)~Similarly, in steps $i=3,4,\dots,9$ we apply the
conditional phase shift $|a\rangle_{i}|b\rangle_{j}\rightarrow e^{i\pi
ab/2^{i-j}}|a\rangle_{i}|b\rangle_{j}$ $(a,b=0,1)$, that is,
\begin{equation}
\widehat{B}_{ij}\equiv\left(
\begin{array}
[c]{cccc}%
1 & 0 & 0 & 0\\
0 & 1 & 0 & 0\\
0 & 0 & 1 & 0\\
0 & 0 & 0 & e^{i\pi/2^{i-j}}%
\end{array}
\right)  =\widehat{B}(\pi/2^{i-j}),
\end{equation}
to each pair of qubits $(i,j)$, $j=1,2,..,i-1$, and then apply the
transformation $\widehat{A}_{i}\equiv\widehat{A}$ to the $i$th qubit. Note that after
the $i$th step the first $i$ qubits store the Fourier transform of the $i$
most significant bits of $x$. Hence, after the 9th step the Fourier transform
is completed:
\begin{equation}
\widehat{F}_{2^{9}}=(\widehat{A}_{9}\widehat{B}_{98}\cdots\widehat{B}_{91})\cdots(\widehat{A}%
_{3}\widehat{B}_{32}\widehat{B}_{31})(\widehat{A}_{2}\widehat{B}_{21})(\widehat{A}_{1}).
\label{DFTAB}%
\end{equation}
Performing these operations implies also application of state exchanges
whenever one needs to transfer the states of atoms $i$ and $j$ to a common
unit cell for processing. A list of control commands ($\delta C_{k}$
sequences) corresponding to Eqs.~(\ref{DFTrev}) and (\ref{DFTAB}) can be
written straightforwardly.

\section{Conclusion}
Non-holonomic control is very relevant in the context of quantum computation, since it allows one to perform any unitary evolution of the system chosen as a computer, or, in other terms, it is a way to implement any quantum gate. Moreover, direct control over the entire system is not always necessary in order to perform computations on the information it contains : indeed, a completely controlled quantum device can be constructed as the smart arrangement of universally controlled unit cells. Two examples of such architectures have been provided in this paper, which are built in accordance with the computational task they are supposed to carry out. Moreover, as an example, a toy device has been considered, which is able to perform a concrete computation on 9 qubits.

\appendix

\section{Toffoli gate} \label{appendix-a}

The Toffoli-gate transformation is the unitary transformation on three qubits,
\begin{equation}
\widehat U_{Toff}|x_2\rangle|x_1\rangle|x_0\rangle
 =|x_2\rangle|x_1\rangle|x_0\,{\rm XOR}\,(x_1\,{\rm AND}\,x_2)\rangle,
\end{equation}
which corresponds to the three-bit classical logic gate,
\begin{eqnarray}
x_2 &\rightarrow& x_2'=x_2 \nonumber\\
x_1 &\rightarrow& x_1'=x_1 \nonumber\\
x_0 &\rightarrow& x_0'=x_0\,{\rm XOR}\,(x_1\,{\rm AND}\,x_2),
\end{eqnarray}
introduced by Toffoli as a universal gate for classical reversible computation.
It acts as a permutation of the computational basis states,
$|x\rangle \equiv |x_2\rangle|x_1\rangle|x_0\rangle$,
$x\equiv\sum_{r=0}^2 x_r 2^r=0,1,\dots,7$, given by the unitary matrix
\begin{equation}
\widehat U_{Toff} =\left(
\begin{array}{cccccccc}
1&0&0&0&0&0&0&0\\
0&1&0&0&0&0&0&0\\
0&0&1&0&0&0&0&0\\
0&0&0&1&0&0&0&0\\
0&0&0&0&1&0&0&0\\
0&0&0&0&0&1&0&0\\
0&0&0&0&0&0&0&1\\
0&0&0&0&0&0&1&0
\end{array}
\right).
\label{UToff}
\end{equation}
This matrix can be presented as
\begin{equation}
\widehat U_{Toff}=\exp(-i\pi\widehat{\cal H}_{Toff}),
\end{equation}
with the (idempotent) Hermitian matrix
\begin{equation}
\widehat {\cal H}_{Toff} =\frac{1}{2}\left(
\begin{array}{cccccccc}
0&0&0&0&0&0&0&0\\
0&0&0&0&0&0&0&0\\
0&0&0&0&0&0&0&0\\
0&0&0&0&0&0&0&0\\
0&0&0&0&0&0&0&0\\
0&0&0&0&0&0&0&0\\
0&0&0&0&0&0&1&-1\\
0&0&0&0&0&0&-1&1
\end{array}
\right).
\label{HToff}
\end{equation}
In our control scheme the Toffoli-gate transformation can be effected
on the unit cell by repeating 8 times the transformation
$\widehat U_{\epsilon/8}\equiv\exp(-i\pi\widehat{\cal H}_{Toff}/8)$
$(\epsilon=\pi)$, which is directly attainable:
$\widehat U(t=64T)=\widehat U_{\epsilon/8}$ (see Fig.~\ref{Fig4}(d)).

\end{document}